# ZnCoO Films by Atomic Layer Deposition – influence of a growth temperature on uniformity of cobalt distribution

M. I. Łukasiewicz[1*], B. Witkowski[2], M. Godlewski[1,2], E. Guziewicz[1], M. Sawicki[1],
W. Paszkowicz[1], E. Łusakowska[1], R. Jakieła[1], T. Krajewski[1],
I.A. Kowalik[1], B.J. Kowalski[1]

[1]*Institute of Physics, Polish Acad. of Sciences, Al. Lotników 32/46, 02-668 Warsaw, Poland*
[2]*Dept. of Mathematics and Natural Sciences College of Science*
*Cardinal S. Wyszyński University, Dewajtis 5, 01-815 Warsaw, Poland*

**Abstract**

We report on the structural, electrical and magnetic properties of ZnCoO thin films grown by Atomic Layer Deposition (ALD) method using reactive organic precursors of zinc and cobalt. As a zinc precursor we applied either dimethylzinc or diethylzinc and cobalt (II) acetyloacetonate as a cobalt precursor. The use of these precursors allowed us the significant reduction of a growth temperature to 300$^o$C and below, which proved to be very important for the growth of uniform films of ZnCoO. Structural, electrical and magnetic properties of the obtained ZnCoO layers will be discussed based on the results of SIMS, SEM, EDS, XRD, AFM, Hall effect and SQUID investigations.

## 1. Introduction

Diluted magnetic semiconductors (DMS) have attracted much attention as potential materials for spintronic devices. ZnTMO, with TM standing for transition metals (Co, Mn, Fe, etc.), is one of the promising DMS materials [1-3]. Theoretical calculations have shown that ferromagnetic state can be obtained at room temperature (RT) in *p*-type ZnMnO [1], but also it may be possible in *n*-type ZnO when doped with cobalt [2, 3]. In fact ferromagnetic ordering at RT in ZnCoO was reported by several groups. However, the origin of the ferromagnetism in ZnCoO is not clear. It is more likely due to metal accumulations and foreign phases, rather than volume properties of samples. Thus, confusing reports on their magnetic properties can be partly related to inclusions of foreign phases and metal accumulations [4,5].

The objective of this investigation is to present and discus the results of structural, magnetic, and electrical measurements made on ZnCoO thin films obtained at low temperature (LT) growth conditions by the Atomic Layer Deposition (ALD) method.
.

## 2. Experimental

The ZnCoO thin films were grown using the ALD technique [6]. We used organic precursors: dimethylzinc and diethylzinc as a zinc precursors, and cobalt (II) acetyloacetonate as a cobalt precursor. Deionized water was used as an oxygen precursor. These highly reactive precursors are sequentially introduced to the growth chamber, so they meet only at a surface of a grown film. The use of these precursors allowed us the significant reduction of a growth temperature to 300 °C and below.

---

[*] mluk@ifpan.edu.pl

Precursors' temperatures were the following: DMZn -30°C, DEZn - RT, Co(acac)$_2$ - 150°C, and water - 30°C. Substrate temperature (we used sapphire, GaAs, Si, glass) ranged between 160°C and 300°C.

## 3. Results

The structural characterization was carried out by X-ray diffraction (XRD) method. XRD spectra for films grown at two different temperatures are shown in Fig.1. The XRD measurements show a polycrystalline nature of ZnCoO layers. In all the cases we observed the diffraction maxima related to (10.0), (00.2) and (10.1) crystallographic orientations. The crystalline structure was significantly better for films grown at lower temperature (160°C) than for the one grown at 300°C. The surface morphology of ZnCoO films was investigated with Atomic Force Microscopy (AFM) and Scanning Electron Microscopy (SEM). RMS value of roughness of the films ranged between 1.28 nm and 22 nm, depending on the film thickness and growth temperature (RMS increases with the thickness but decreasing with increasing temperature).

SEM images taken at relatively high magnification are shown in Fig.2 for two growth temperatures 300°C (a) and 160°C (c). SEM images show flake-like micro-structure of ZnCoO films grown with a few percent Co fraction.

In Fig.2 (a,c) we show islands formed on a surface of the studied films (such islands appear very sporadic). The chemical composition of these islands was determined by using the Energy Dispersive Spectroscopy (EDS) method. Results of EDS measurements are shown in Fig.2. (b) and (d). In the case of a higher growth temperature these islands consist of a CoO foreign phase, which is not the case of the samples grown at lower temperature. Maps of the EDS spectra taken from flat surface area indicated fairly in-plane uniform Co distribution for both types of the ZnCoO films.

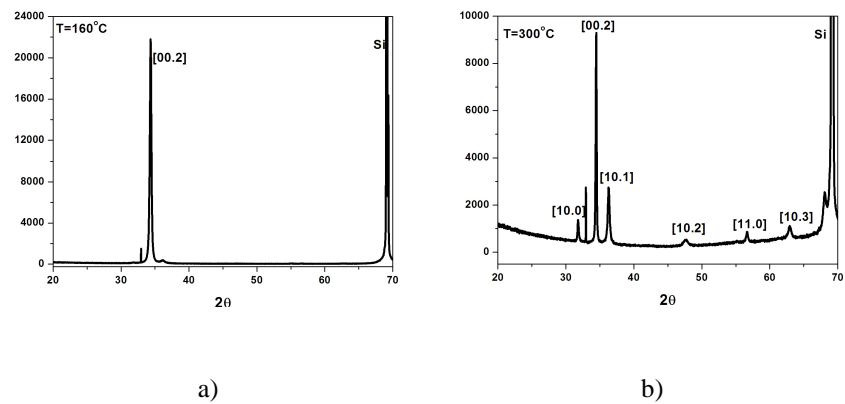

a) b)

Fig. 1. X-ray diffraction (XRD) spectra for ZnCoO films grown at different two temperatures: (a) 160°C, (b) 300°C.

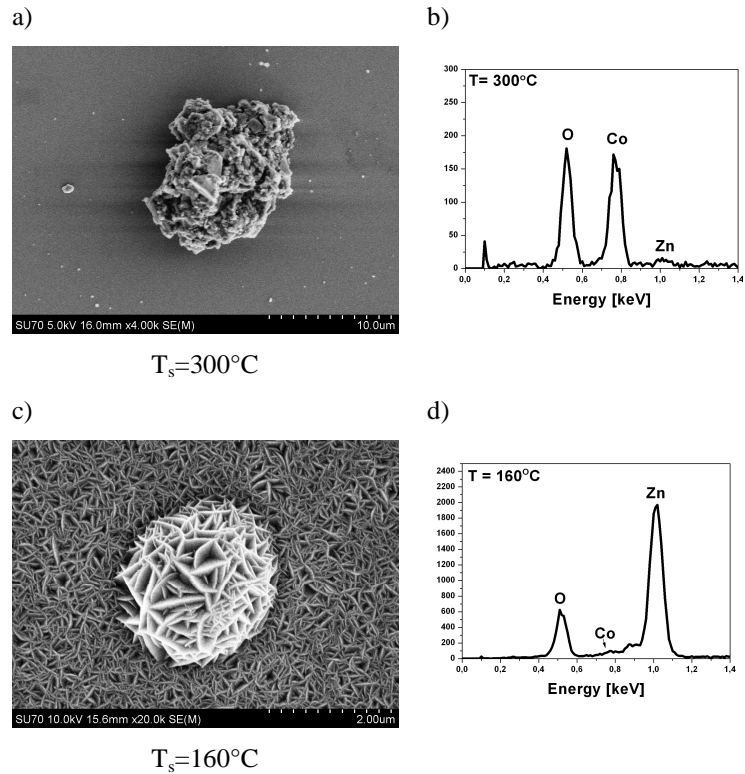

Fig. 2. SEM images (left) and EDS (right) spectra for ZnCoO films grown at two different temperatures: (a) 160°C, (b) 300°C.

Depth uniformity of Co concentration in ZnCoO films was studied with secondary ion mass spectroscopy (SIMS). SIMS investigations showed high homogeneity of the ZnCoO layers, when grown with the optimized ALD growth parameters. To optimize these parameters we tested several sequences of elementary ALD cycles and times of pulses of cobalt precursor. We obtained the best results for the following sequences of the ALD cycles - eighty cycles to form layers of ZnO followed by five cycles to form layers of CoO. Films grown with thinner ZnO "spacers" were less uniform and showed worse crystallographic structure.

Figs. 2 (a) and (c) illustrate difficulties in obtaining more ordered films. We found two effects related to an increasing temperature of the process. First, Co fraction in ZnCoO increases (from 1.4 % to 5.5 atomic %) in the samples grown at higher temperature, even though precursor pulses were the same. Second, the growth mode depends on a growth temperature. It is more 3D-like for samples grown at low temperature (see Fig. 2 (c)), with columns (flakes-like) of mixed orientation, as deduced from the XRD investigations. To obtain films with a better crystallographic structure it was necessary to reduce Co fraction to say below 5%, lower growth temperature and to introduce relatively thick ZnO "spacers" in the ALD growth cycles.

The Hall measurements conducted on the samples of ZnCoO layers showed that the electron concentration varied between $10^{15}$ cm$^{-3}$ and $10^{19}$ cm$^{-3}$, depending on the growth temperature, Co fraction and time of application of Co-precursor (Fig.3). Important observation is that we can achieve high n doping in ZnCoO films. On the other hand we observed a surprising anti-correlation between length of Co precursor pulses and carrier concentration and mobility (see Fig. 3).

After setting optimal conditions for ZnCoO deposition we performed magnetization investigations using SQUID magnetometer. Films of ZnCoO with homogeneous Co distribution, with Co fraction below 5%, and these which were grown at LT show paramagnetic properties at RT. Nonuniform films show multi-phase signals in SQUID measurements.

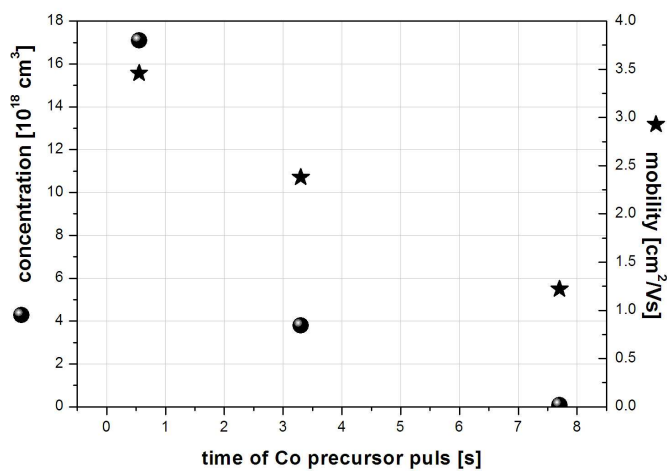

Fig. 3. Anti-correlation between time of Co precursor pulses and sample conductivity (electron concentration and mobility).

## 4. Conclusions

It is important to underline that ZnCoO turned out to be by far more difficult to be grown as uniform films (as compared e.g. to ZnMnO). Anyway, by optimizing sequences of the ALD cycles and the growth temperature we have grown uniform ZnCoO films showing only paramagnetic response in the SQUID measurements. We also observed that high n-type doping is possible in ZnCoO.

**Acknowledgements:**

This work was supported by FunDMS ERC Advanced Grant Research.